# EFFECT OF TRANSVERSE ELECTRON VELOCITIES ON THE LONGITUDINAL COOLING FORCE IN THE FERMILAB ELECTRON COOLER

Andrei Khilkevich[#], BSU, Minsk, Belarus
Lionel R. Prost, Alexander V. Shemyakin, Fermilab, Batavia, IL 60510, USA

*Abstract*

In Fermilab's electron cooler, a 0.1A, 4.3MeV DC electron beam propagates through the 20 m cooling section, which is immersed in a weak longitudinal magnetic field. A proper adjustment of 200 dipole coils, installed in the cooling section for correction of the magnetic field imperfections, can create a helix-like trajectory with the wavelength of 1-10 m. The longitudinal cooling force is measured in the presence of such helixes at different wavelengths and amplitudes. The results are compared with a model calculating the cooling force as a sum of collisions with small impact parameters, where the helical nature of the coherent angle is ignored, and far collisions, where the effect of the coherent motion is neglected. A qualitative agreement is found.

## INTRODUCTION

The Recycler electron cooler [1] will be decommissioned after the end of the Tevatron Run II in October 2011. The possibility of re-using its components for implementation of electron cooling at BNL to improve the life time at RHIC during a low energy campaign under discussion [2] is presently being studied. One of the issues for electron cooling in RHIC is a beam loss associated with recombination of ions with electrons from the cooling beam. Reference [3] proposed to suppress the recombination by introducing undulator fields in the cooling section. A helical trajectory of the electron beam centroid introduces a coherent angle $\alpha_h$, which can significantly reduce recombination while affecting only modestly the cooling force. It happens if the helix period $\lambda$ is short enough

$$\rho_h \equiv \alpha_h \frac{\lambda}{2\pi} \ll \rho_{\max}, \qquad (1)$$

where $\rho_{\max}$ is the maximum impact parameter, and the contribution of collisions with impact parameters $\rho \gg \rho_h$ ("far" collisions) is dominant.

At Fermilab's Recycler, experiments were conducted in order to try to quantify the effect on cooling of a helical trajectory as proposed in [3]. However, the cooler has two important differences with the scheme of [3]. First, the Recycler cooler employs a 105 G longitudinal magnetic field to transport the electron beam through the cooling section. But, because of the low value of the field, the main contribution to the cooling force comes from collisions where the effect of the magnetic field is negligible. The model of interaction with a free electron gas [4] was successfully used to fit experimental results when characterizing the electron cooling force in the Recycler [5].

Second, the installed equipment doesn't allow creating a helical trajectory with a wavelength shorter than $\lambda = 1$m, which is an order of magnitude larger than considered in [3]. Correspondingly, the contribution of small impact parameters ("near" collisions region) is significant or even dominating. Therefore, to show the effect of far collisions, we compare the cooling force at a constant angle and different wavelengths.

## THE COOLING FORCE IN THE PRESENCE OF A HELICAL MOTION

For the purpose of discussing the experimental data, let's consider the following case:

- The rms transverse and longitudinal electron velocities in the beam frame are as measured previously [5], $\sigma_{x,y} \sim 4 \times 10^5$ m/s and $\sigma_z \sim 2 \times 10^4$ m/s (corresponding to the electron angle of $\alpha_e \sim 0.15$ mrad and electron energy spread of ~300eV in the lab frame);
- The longitudinal cooling force is measured by the voltage jump method (see description in [6]) and the electron beam energy jump is chosen to be 2 KeV. It results in a difference of longitudinal velocities $\Delta V_z \sim 1.2 \times 10^5$ m/s in the beam frame, which is significantly larger than other antiproton velocity components;
- The beam centroid undergoes a helical motion due to either Larmor rotation in the longitudinal magnetic field of 105 G ($\lambda = \lambda_B = 9.6$ m) or dipole correctors set appropriately.

Following the logic of Ref. [7], near collisions of an antiproton with electrons at $\rho < \rho_h$ do not "feel" that the coherent angle comes in part from the helical motion, while, far collisions with $\rho \gg \rho_{osc}$, are insensitive to the coherent angle. Correspondingly, the cooling force can be estimated as a sum of near and far collisions with the impact parameter transition from "near" to "far" collisions at $\rho = k_d \cdot \rho_h$ with an empirical coefficient $k_d \sim 1$.

Using the formalism of Ref. [7], this leads to the following set of equations

$$F_Z = 4\pi n_e m_e r_e^2 c^4 \eta \cdot$$

$$\left\{ \iiint \frac{\ln\left(\frac{k_d \rho_h}{\rho_{min}}\right)(v_z + \Delta V_z) f(v_e)}{\left[v_x^2 + v_y^2 + (v_z + \Delta V_z)^2\right]^{3/2}} d^3 v_e + \right.$$

$$\left. \iiint \frac{\ln\left(\frac{\rho_{max}}{k_d \rho_h}\right)(v_z + \Delta V_z) \tilde{f}(v_e)}{\left[v_x^2 + v_y^2 + (v_z + \Delta V_z)^2\right]^{3/2}} \right\} \quad (2)$$

where

$$f(v_e) = \frac{1}{(2\pi)^{3/2} \sigma_x \sigma_y \sigma_z} e^{-\left(\frac{v_x^2}{2\sigma_x^2} + \frac{v_y^2}{2\sigma_y^2} + \frac{v_z^2}{2\sigma_z^2}\right)} \quad (3)$$

$$\tilde{f}(v_e) = \frac{1}{(2\pi)^{3/2} \sigma_x \sigma_y \sigma_z} e^{-\left(\frac{v_x^2}{2\sigma_x^2} + \frac{(v_y - U)^2}{2\sigma_y^2} + \frac{v_z^2}{2\sigma_z^2}\right)} \quad (4)$$

$$\rho_{min} = \frac{r_e c^2}{v_x^2 + v_y^2 + (v_z + \Delta V_z)^2}, \quad (5)$$

$n_e$ - electron density in the beam rest frame; $m_e$ - electron mass; $r_e$ - electron classical radius; $\eta = 0.006$ is the ratio of the ring length occupied by the cooling section to the Recycler ring circumference; $U$ is an additional coherent transverse velocity, coming from the helical motion of the beam; $\Delta V_z$ - electron velocity increase due to the voltage jump; $\sigma_x$, $\sigma_y$, $\sigma_z$ - rms velocity spread on X, Y and Z axis respectively.

## GENERATION OF A HELIX

The cooling section consists of 10 identical 2 meter-long modules. Each module is equipped with 10 pairs (X and Y) dipole correction coils and a pair of capacitive pickups (BPMs). All 200 correctors underwent a beam-based calibration, which is believed to be correct within several percents.

Mechanical positions of the modules are drifting with time and variations of the cooling section temperature. The drift results in a degradation of the electron trajectory and a corresponding decrease of the cooling force due to an increase of the effective electron rms angle $\alpha_e$. The beam trajectory cannot be measured inside modules, and the only way to estimate the change of $\alpha_e$ is from the cooling measurements. Note that the mechanical positions of BPMs are drifting at the same time. This is corrected within the signal processing software according to measurements with the same BPMs of the antiproton trajectory in the cooling section which is known to be a straight line since the transverse magnetic field is too weak to introduce any measurable deviation.

The helical trajectory was generated either by a set of correctors just upstream of the cooling section (by inducing a simple Larmor rotation) or by all 200 correctors of the cooling section. Corrector currents' calculation was based on the program developed in [8]. For simplification purposes, the wavelength of the helix was chosen to be a multiple of the module length. The helix was concentric with the antiproton trajectory. The measured trajectories were reasonably close to helixes (Fig. 1).

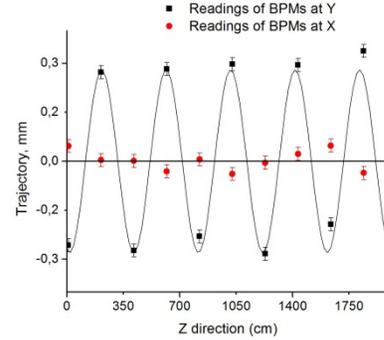

Figure 1. Example of a helix with $\lambda_h$ = 4m, $\alpha_h$ = 0.4mrad. Data points are BPM readings, and the solid cosine-like line represents the goal Y position. Ideally, the data points for X (red circles) should be exactly zero.

## EXPERIMENTAL RESULTS

Results of the measurements of the longitudinal cooling force for various angles and periods of the helix at the electron beam current of $I_e$ = 0.1A are presented in Fig. 2. The solid curve shows a calculation that does not take into account the contribution of far collisions (Eq. (2)). The value of the current density at the beam center, $j_{o\_cs}$= 0.96 A/cm$^2$, is extrapolated from the electron gun simulation and the ratio of magnetic fields at the cathode and in the cooling section. Then, the angular spread $\alpha_e$ was adjusted to fit the value of the force measured without excitation of a helix *i.e.* for the data set labelled $\lambda_h$ = 9.6m and found to be $\alpha_e$ = 0.14 mrad.

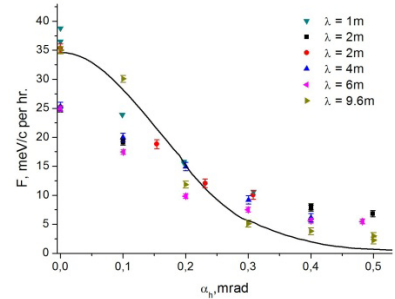

Figures 2. The cooling force as a function of the helix angle at several wavelengths. Error bars indicate a statistical error in fitting the data.

The data sets for different wavelengths presented in Fig.2 were measured over several months, and the results for zero helical angle have a large scatter due to the mechanical drifts mentioned above. However, for large angles $\alpha_h^2 >> \alpha_e^2$, contribution of this scatter to the cooling force should be insignificant, and the data set can be used to deduce the dependence of the cooling force on the helix period as shown in Fig. 3. Fitting Eq. (2) to the data set with $\alpha_h = 0.4$ mrad (red solid line) gave $k_d = 1.3$. The other solid lines on the plot are calculated using Eq. (2) with the same $k_d$.

Note that if the near collisions were the only contribution to the cooling force, the data in Fig. 3 would lay on straight lines. A monotonic growth with a decrease of the wavelength is indicative of the contribution of far collisions, which dominates at $\lambda_h = 1$ m.

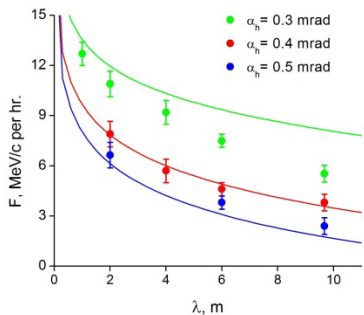

Figure 3. The electron cooling force as a function of the helix wavelength for $\alpha_h = 0.3$, 0.4, and 0.5 mrad.

While the calculated curves are qualitatively similar to the trend in the data, there is a significant quantitative difference. Part of the disagreement can be attributed to the implicit assumption in the model that the helix radius is small in comparison with the beam size, $R_{beam} >> \rho_h$. The highest value of the helix radius $\rho_h$ is 0.75 mm (for $\alpha_h = 0.5$ mrad). However, the simulated current density distribution in the cooling section is close to parabolic [5]

$$j_e(r) \approx j_{0\_CS} \cdot \left(1 - r^2/a^2\right) \quad (6)$$

with $a = 2.9$ mm, which is of the same order as $\rho_h$. The most obvious consequence is that the antiprotons now sample the electron beam at $\rho_h$ from its center, where the current density is lower (and, generally speaking, $\alpha_e$ is larger). This effect can change the force by up to ~10%.

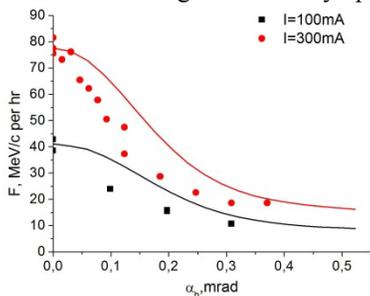

Figure 4. The electron cooling force as a function of the helix angle for two beam currents at $\lambda_h = 1$ m.

Comparison of the cooling force measured for two values of the electron beam current is shown in Fig. 4. The solid lines are a calculation with Eq. (2) for $k_d = 1.3$, $j_{0\_cs} = 1.6$ A/cm$^2$ for $I_e = 0.3$A. The values of $\alpha_e$ are such that the cooling force calculated for $\alpha_h = 0$ is the same as the one measured in each case (0.12mrad and 0.13 mrad for $I_e = 0.1$ A and 0.3A, respectively). While a decrease of the cooling force due to a coherent angle can obviously be compensated by an increase of the beam current, the net effect on the compensation rate requires a more detailed analysis.

The results presented above also allow drawing a conclusion related to the degradation of the cooling efficiency during regular operation of the Recycler electron cooler over time. Trajectory perturbations caused by mechanical drifts of the cooling section degrade the cooling efficiency in a manner similar to that of the helical trajectories introduced in these studies. Hence, correction of the long-wave perturbations on the scale of the whole-section ($\lambda_h \sim 10$m) is likely to have a larger impact than adjustments within a single module ($\lambda_h \sim 1$m).

## CONCLUSION

1. The longitudinal cooling force was measured in the presence of a coherent helical motion of the electron beam with different amplitudes and wavelengths. The observed growth of the cooling force with a decrease of the helix wavelength indicates that collisions with impact parameters larger than the helix radius effectively contribute to the cooling force
2. The data were compared with a model calculating the cooling force as a sum of near collisions, where the helical nature of the coherent angle is ignored, and far collisions, where the effect of the coherent motion is neglected. With one fitted parameter characterizing the boundary between the two regions, the calculation agrees with the data within ~30%.
3. The measurements support the feasibility of the scheme proposed for suppression of recombination in the RHIC cooler.